**Non-Fermi-liquid behavior in nearly ferromagnetic metallic SrIrO$_3$ single crystals**

G. Cao[1], V. Durairaj[1], S. Chikara[1], L. E. DeLong[1], S. Parkin[2], and P. Schlottmann[3]

[1]Department of Physics and Astronomy, University of Kentucky, Lexington, KY 40506

[2]Department of Chemistry, University of Kentucky, Lexington, KY 40506

[3]Physics Department, Florida State University, Tallahassee, FL 32306

We report transport and thermodynamic properties of single-crystal SrIrO$_3$ as a function of temperature T and applied magnetic field H. We find that SrIrO$_3$ is a non-Fermi-liquid metal near a ferromagnetic instability, as characterized by the following properties: (1) small ordered moment but no evidence for long-range order down to 1.7 K; (2) strongly enhanced magnetic susceptibility that diverges as $T^{-\gamma}$ at low temperatures with $1/2 < \gamma < 1$, depending on the applied field; (3) heat capacity C(T,H) ~ -Tlog T that is readily amplified by low applied fields; (4) a strikingly large Wilson ratio at T< 4K; and (5) a $T^{3/2}$-dependence of electrical resistivity over the range 1.7 < T < 120 K. A phase diagram based on the data implies SrIrO$_3$ is a rare example of a stoichiometric oxide compound that exhibits non-Fermi-liquid behavior near a quantum critical point (T = 0 and $\mu_o$H = 0.23 T).

The discoveries of exotic ground states (p-wave superconductivity, non-Fermi liquid (NFL)) [1, 2] in layered ruthenates have inspired extensive investigations on 4d and 5d materials. Typified by their extended 5d orbitals, it is commonly expected that iridates should be more metallic and less magnetic than their 3d, 4d and 4f counterparts, because of the broader 5d-bandwidth and the weaker exchange interaction $U$, so that $Ug(E_F) < 1$, where $g(E_F)$ is the density of states at the Fermi energy. However, in marked contrast to these expectations, most of the known iridates, such as layered $BaIrO_3$ [3-7] and $Sr_{n+1}Ir_nO_{3n+1}$ (n = 1 and 2) [8-14], are insulators exhibiting weak ferromagnetism [15]. On the other hand, the layered 4d ruthenate analogs ($BaRuO_3$, $Sr_2RuO_4$ and $Sr_3Ru_2O_7$) are metallic or even superconducting. Although the layered iridates order at relatively high temperatures ($T_c$ = 175, 240, and 285 K for $BaIrO_3$, $Sr_2IrO_4$ and $Sr_3Ir_2O_7$, respectively), they attain only a small fraction of the expected ordered moment ($\mu_{or}$ = 0.03, 0.14 and 0.037 $\mu_B$/Ir, respectively) [6, 12, 14]. The iridates exhibit strong phase transition signatures in magnetization but none of their $T_c$'s and resistivities are very sensitive to high magnetic fields [6, 12, 14]. Although a metallic state does not commonly occur in the iridates, the unusual circumstances almost guarantee that it will exhibit extraordinary properties when it does occur. In this paper, we report anomalous transport and thermodynamic properties of single-crystal $SrIrO_3$, which we find is a NFL metal with a ferromagnetic instability extrapolated to zero-temperature at an applied magnetic field $\mu_oH = 0.23$ T.

There are several examples of intriguing quantum phenomena occurring in itinerant-electron materials that are on the borderline between ferromagnetism and paramagnetism [16-18], e.g. p-wave superconductivity in $Sr_2RuO_4$ [1], superconductivity and ferromagnetism in ZrZn$_2$ [19] and URhGe [20], a ferromagnetic quantum critical point (QCP) in MnSi under pressure [21], a metamagnetic transition with QCP end-point tuned by a magnetic field in Sr$_3$Ru$_2$O$_7$ [2], and QCP with anomalous ferromagnetism in Sr$_4$Ru$_3$O$_{10}$ [22]. In numerous Ce, Yb and U compounds similar phenomena associated with a QCP, but with *antiferromagnetic* spin-fluctuations have been found [24], leading to the breakdown of Fermi-liquid behavior, including a divergent specific heat [C/T~ -logT] and unusual power laws in resistivity ρ and magnetic susceptibility χ at low temperatures [24, 25]. The QCP can be tuned by "control parameters" such as composition, pressure, magnetic field, etc. [23-25]. To our knowledge SrIrO$_3$ is the first *stoichiometric* metallic 5d system with a nearby QCP that can be readily tuned with *very modest magnetic fields*; such a rare combination makes SrIrO$_3$ a unique and desirable model system for studies of quantum criticality both experimentally and theoretically.

It is now qualitatively understood that the extended 5d orbital significantly enhances crystalline electric field splittings causing a partial breakdown of Hund's rule, leading to a low spin state, S = 1/2 for Ir$^{4+}$ (5d$^5$) ions, and a *d-p* hybridization between the transition metal and the oxygen octahedron enclosing it. These interactions generate strong electron-lattice coupling that can alter and distort the metal-oxygen bond lengths and angles, lifting the degeneracies of 5d-orbitals and precipitating orbital ordering.

Single crystals were grown using flux techniques described elsewhere [7]. The crystal structure was determined from a small fragment (0.05×0.05×0.05 mm$^3$) using Mo *Kα* radiation and a Nonius Kappa CCD single-crystal diffractometer. Heat capacity measurements were performed with a Quantum Design PPMS that utilizes a thermal-relaxation calorimeter operating in fields up to 9 T. Magnetic and transport properties were measured using a Quantum Design MPMS 7T LX SQUID Magnetometer equipped with a Linear Research Model 700 AC bridge for transport measurements. We define the susceptibility as $\chi \equiv M/H$ in terms of the DC magnetization M(T,H).

The crystal structure of SrIrO$_3$ is a monoclinic distortion of the hexagonal BaTiO$_3$ or 6H structure [26] with space group C2/c (No.15) having lattice parameters $a$ = 5.604 Å, $b$ = 9.618 Å, $c$ = 14.170 Å and $\beta$ = 93.26°. It features a distorted six-layer hexagonal structure that consists of close-packed Sr-O layers stacked perpendicular to the c-axis in the sequence hcchcc, where h and c refer to hexagonal and cubic stacking. Ir-O octahedra share common corners across a c-layer and common faces across an h-layer, forming pairs of face-shared octahedra that are joined by common corners to a plane of corner-sharing octahedra.

Fig. 1a shows the DC magnetic susceptibility $\chi$ as a function of $T$ at $\mu_oH$ = 0.5 T for **H** ∥ **c**-axis ($\chi_c$) and **H** ⊥ **c**-axis ($\chi_{ab}$). Both $\chi_c$ and $\chi_{ab}$ are clearly temperature-dependent and large, particularly for $T$ < 15 K. In contrast, IrO$_2$ is a Pauli paramagnet with $\chi$ (~$10^{-4}$ emu/mole-Ir), i.e. $\chi$ is temperature-independent and much smaller than that of SrIrO$_3$ (>$10^{-3}$ emu/mole-Ir). While the large values of $\chi_c$ and $\chi_{ab}$ for SrIrO$_3$ are most likely the consequence of a significant exchange enhancement, $Ug(E_F)$, the sharp rise below 15 K suggests the proximity to a ferromagnetic instability. (Note that $\chi_c$ surpasses $\chi_{ab}$ at $T$ <5 K.) Indeed, the isothermal magnetization M at $T$ = 1.7 K is already saturated at $\mu_oH$ ~3 T, as seen in the inset to Fig. 1a. On the other hand, the corresponding ordered moment $\mu_{or}$ is less than 3% of that expected (1 $\mu_B$/Ir) for an S = 1/2 system and decreases with increasing T, which is indicative of a nearby Stoner instability.

The reciprocal susceptibilities $\chi_c^{-1}$ and $\chi_{ab}^{-1}$ display linear $T$-dependences, consistent with a Curie-Weiss behavior for $T > 120$ K, as shown in Fig.1b. However, the Curie-Weiss fits of the data yield effective moments and Curie-Weiss temperatures that are much too large to be physically meaningful. This behavior is similar to the observed for the exchange-enhanced paramagnet $SrRhO_3$ [27, 28] and the metallic weak-ferromagnets $ZrZn_2$ [29] and $Sc_3In$ [30]. Moreover, for $1.7 < T < 15$ K, $\chi_c^{-1}$ and $\chi_{ab}^{-1}$ follow non-standard power laws that range from $T^{1/2}$ for B < 0.3 T to linear-$T$ for B > 0.8 T, as shown in Fig. 1c. This high sensitivity of the temperature exponent to low applied magnetic fields again suggests the rapid approach to a ferromagnetic instability.

The low temperature specific heat C($T$,H) data acquired over $1.8 < T < 24$ K and $\mu_o H < 8$ T offer important insights into the low energy excitations of $SrIrO_3$. For $T > 12$ K, the specific heat is well described by $C(T) = \gamma T + \beta T^3$ with $\gamma = 1.50$ mJ/mole K$^2$ and $\beta = 0.28$ mJ/mole K$^4$, suggesting that only electronic and phonon contributions are significant in this temperature range (data not shown). The small $\gamma$-value implies that there is essentially no mass enhancement and from the $\beta$-value we obtain a Debye temperature of 326 K.

The heat capacity data exhibit intriguing temperature and field dependences below 13 K, as shown in Fig.2a, where $C/T$ vs $T$ is plotted for $1.8 < T < 12.8$ K. A broad shoulder is observed near 4.5 K, which weakens with increasing field and eventually vanishes at $\mu_o H > 3$ T. The field dependence suggests a magnetic mechanism, but $\chi(T,H)$ shows no corresponding transition. This broad peak in $C/T$ is followed at lower temperatures by a pronounced log($T$) dependence, characteristic of NFL systems [19], and suggests a vanishing Fermi temperature ($T_F \rightarrow 0$) and a divergent quasi-particle effective mass ($m^*/m \to \infty$). It is clear that the amplitude of the logarithmic term rapidly grows with increasing field until $\mu_o H = 1.1$ T, where it becomes weaker, and eventually vanishes for $\mu_o H > 2$T. Noticeably, $C/T$ below 6 K is essentially constant for $\mu_o H = 3$ T and then slowly drops off with decreasing $T$ and increasing H, which indicates the removal of magnetic entropy and the recovery of Fermi-liquid behavior. This crossover is illustrated in Fig. 2a, which also presents a plot of C vs $T^{3/2}$ (right and upper scales) at $\mu_o H = 8$ T. Hence, due to a magnetic field of 8 T, $C(T,H)$ evolves from the NFL $T\log(T)$ behavior to a $T^{3/2}$ power law expected for ferromagnetic magnons out of an ordered state.

The magnetic contribution to the heat capacity, $\Delta C$, at low temperatures is obtained by subtracting the electronic ($\gamma T$) and phonon ($\beta T^3$) contributions that dominate $C(T)$ in the range $10 < T < 24$ K. The plot of $\Delta C/T$ vs. $T$ shown in Fig. 2b emphasizes the logarithmic behavior of $\Delta C/T$ for B = 0 and 1.1 T for $T < 10$ K. The strong enhancement of $\Delta C/T$ with only weak applied fields < 1.5 T reflects the growth of quantum critical fluctuations near a ferromagnetic instability. The strong competition between a ferromagnetic state and spin fluctuations can be inferred from the intermingling of the log T-dependence and the hump in $\Delta C/T$ located near 4.5 K, which, for $\mu_o H = 3$ and 5T, broadens as the low-temperature, singular behavior of $\Delta C/T$ disappears.

The detailed field dependence of $C/T$ reveals two interesting features shown for representative isothermals in Fig. 2c. (1) $C/T$ peaks at a critical field $H_c$ that separates a regime for $H < H_c$ where $C/T \sim -\log(T)$ increases with H, from the complementary regime for $H > H_c$ where the $\log(T)$ dependence weakens and eventually disappears. The peak fades and $C/T$ becomes much less field dependent for $T > 4$ K. On the basis of the $C(T,H)$ data an H-T phase diagram (Fig. 3a) can be constructed to reveal a linear increase of $H_c$ with temperature that can be extrapolated to $T = 0$ K to locate the QCP at $\mu_oH_c = 0.23$ T. (2) All $C/T$ curves converge at $\mu_oH = 3$ T (indicated by an arrow in Fig. 2c). This is unlikely to be a coincidence, as an applied field $\mu_oH = 3$ T clearly renders $C/T$ temperature independent (see Fig.2a), indicating a possible crossover to Fermi liquid behavior.

It is also worth mentioning that the Wilson ratio, $R_W \equiv 3\pi^2 k_B^2 \chi / \mu_B^2 \gamma$, is 74.96 at $T = 1.8$ K and shows weak temperature dependence below 4 K, as shown in Fig.3b. This strikingly large $R_W$ is clearly the consequence of incipient ferromagnetism, and is far beyond the values (e.g., $R_W \sim 1$-$6$) typical of heavy Fermi liquids and exchange-enhanced paramagnets such as Pd [31]. But $R_W$ drops rapidly at $T > 4$ K where the $C/T$ peak fades (see Fig.3a), reaffirming the crossover to Fermi liquid behavior (see Fig.3b).

The presence of the quantum critical fluctuations is further corroborated by the temperature dependence of the **c**-axis $\rho_c$ and **ab**-plane $\rho_{ab}$ resistivities as a function of $T$, shown in Fig. 4a. The residual resistivty $\rho_o$ is 2.2 $\mu\Omega$ cm and 0.66 m$\Omega$ cm for $\rho_{ab}$ and $\rho_c$, respectively, and the residual resistance ratio RRR $\approx 3$. An interesting feature is that $\rho_c$ and $\rho_{ab}$ exhibit a $T^{3/2}$ law over a wide temperature range up to 120 K, which is particularly strong for $\rho_c$, as shown in Fig. 4b, where $\rho_c$ vs $T^2$ (upper scale) is also shown for comparison. At $\mu_oH \geq 5$ T, the temperature dependence of $\rho$ changes from $T^{3/2}$ to $T^2$ (see inset), suggesting a recovery of Fermi liquid behavior. It is remarkable that $\rho$ exhibits a large anisotropy ($\rho_c/\rho_{ab} \sim 300$) that is essentially temperature independent, implying quasi-two-dimensional transport, although the magnetic susceptibility is much less anisotropic, suggesting three-dimensional magnetic correlations. There are no indications for long-range orbital order, possibly because the Ir-ions are on sites of low symmetry.

The $T^{3/2}$ and $T^{5/3}$ laws are seen in QCP systems such as MnSi [24], $Sr_3Ru_2O_7$ [2], $Sr_4Ru_3O_{10}$ [23, 32], and some heavy fermion systems [24]. The $T^{5/3}$-dependence is attributed to dominant low-angle electron scattering by low-$q$ spin fluctuations [24], hence weakening the temperature dependence of the resistivity from $T^2$. The power-law $T^{3/2}$ is thought to be associated with effects of diffusive electron motion caused by strong interactions between itinerant electrons and critically damped very-long-wavelength magnons [19]. Electron scattering by static imperfections will render these modes inoperative and a $T^2$-dependence expected in Fermi liquid theory will result. The presence of the $T^{3/2}$ behavior at low temperature is therefore another indication that the crystals studied are of high quality.

All results presented here coherently support a scenario that $SrIrO_3$ is quite close to a QCP and that the observed physical properties are overwhelmingly dominated by strong spin fluctuations. This first study on single-crystal $SrIrO_3$ reveals it to be a *stoichiometric oxide* with unusual sensitivity to low magnetic fields, which makes it an outstanding model system for studies of quantum criticality. Given the phase diagram of Fig. 3a and the association of weak ferromagnetism with triplet-paired superconductivity near a QCP, it is urgent to explore the physical properties of $SrIrO_3$ in the milli-Kelvin range and at high pressures.

**Acknowledgement**: G.C. would like to thank Dr. W. Crummett for useful discussions. This work was supported by NSF through grants DMR-0240813 and 0552267. P.S. is supported by the DOE through grant DE-FG02-98ER45707.

**Captions:**

**Fig.1**. (a) The magnetic susceptibility $\chi$ as a function of temperature at $\mu_oH= 0.5$ T for H || **c**-axis ($\chi_c$) and H$\perp$**c**-axis ($\chi_{ab}$). $\chi$ for polycrystalline IrO$_2$ is also shown for comparison; Inset: Isothermal magnetization M vs. H at T=1.7 K; (b) Reciprocal susceptibility $\chi_c^{-1}$ and $\chi_{ab}^{-1}$ as a function of temperature for $\mu_oH= 0.5$ T; (c) $\chi_c^{-1}$ as a function of $T^{1/2}$ and T (upper scale, for B=0.8 T marked by an arrow). The behavior of $\chi_{ab}^{-1}$ is similar and not shown.

**Fig.2.** (a) The specific heat C divided by temperature, C/T, vs. log T for $\mu_oH=0$, 0.5, 1.1, 3, 5, and 8T; and C vs. $T^{3/2}$ (right and upper scales) for $\mu_oH= 8$ T. (b) $\Delta C/T$ vs. log T (see definition of $\Delta C$ in text) for $\mu_oH= 0$, 1.1, 3, 5 T; (c) C/T vs.H for some representative temperatures.

**Fig.3.** (a) An H-T phase diagram generated based on the data in Fig. 2. The dashed line is a guide to the eye; (b) The Wilson ratio $R_W$ as a function of T. $R_W$ is estimated based on $\chi$ and C/T at $\mu_oH=0.5$ T.

**Fig.4.** (a) The basal plane and c-axis resistivity, $\rho_{ab}$ and $\rho_c$ (right scale) as a function of temperature; (b) $\rho_c$ vs. $T^{3/2}$ and $T^2$ (upper scale) for 1.7<T<120 K at $\mu_oH=0$ T. Inset: $\rho_c$ vs. $T^2$ for 1.7<T<37 K at $\mu_oH=5$ T.

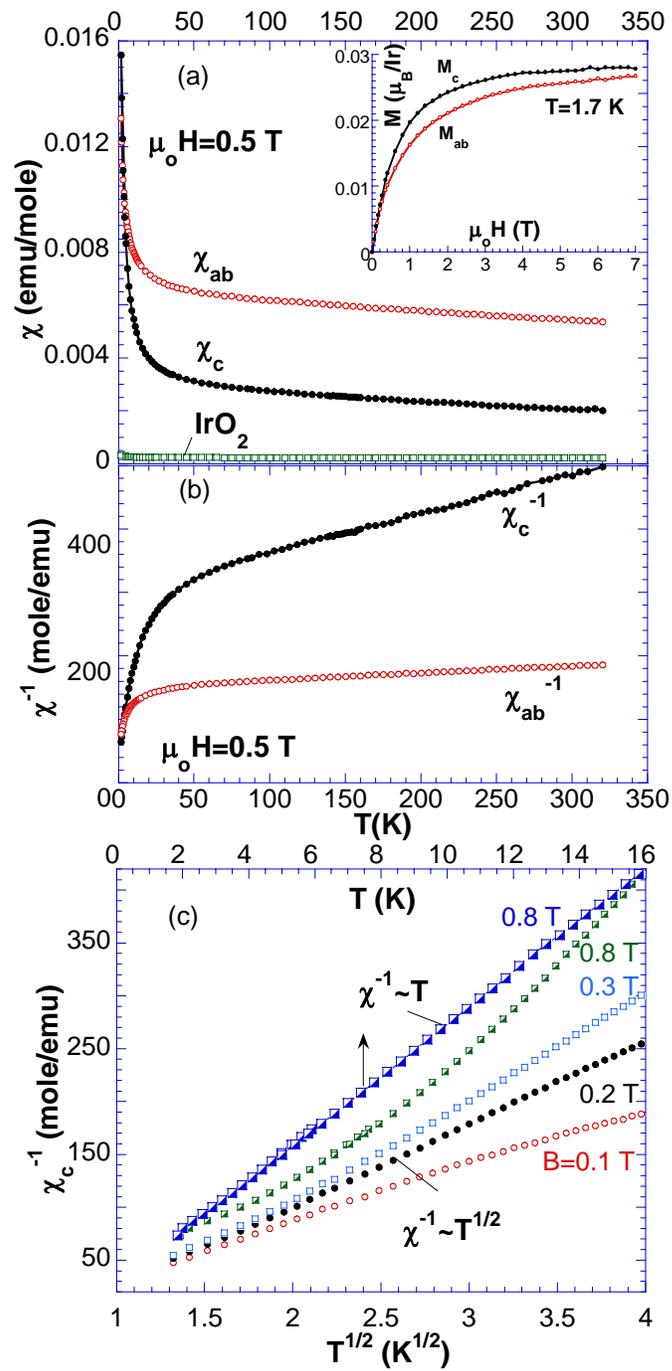

Fig. 1

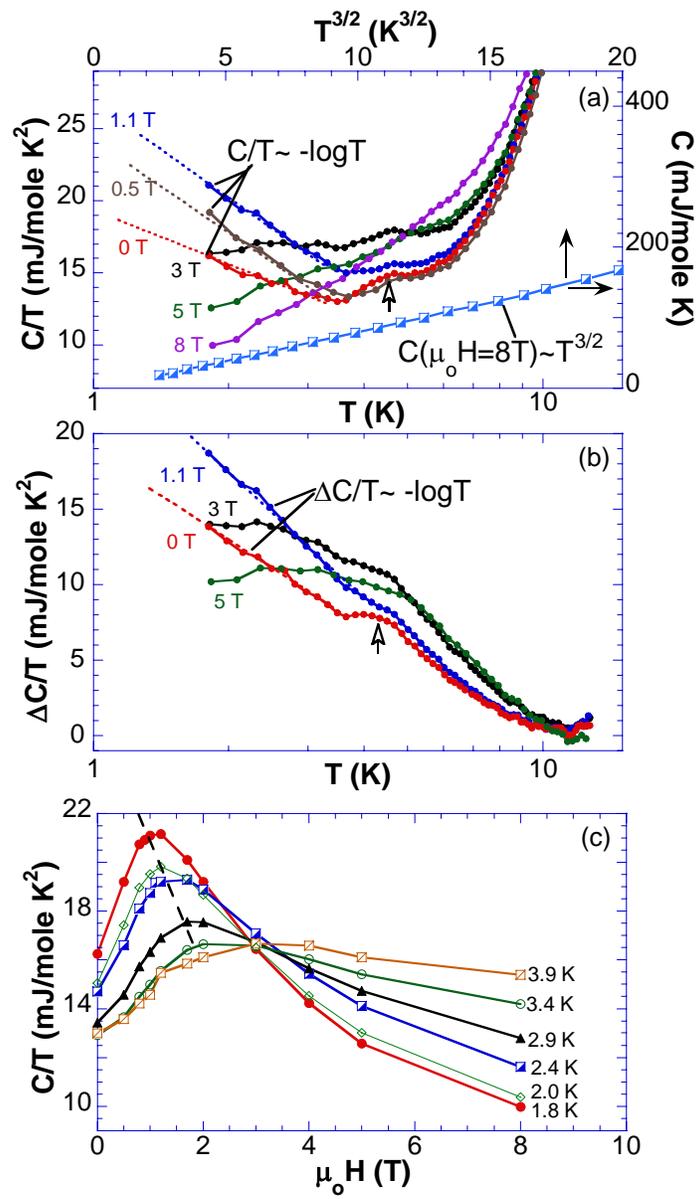

Fig. 2

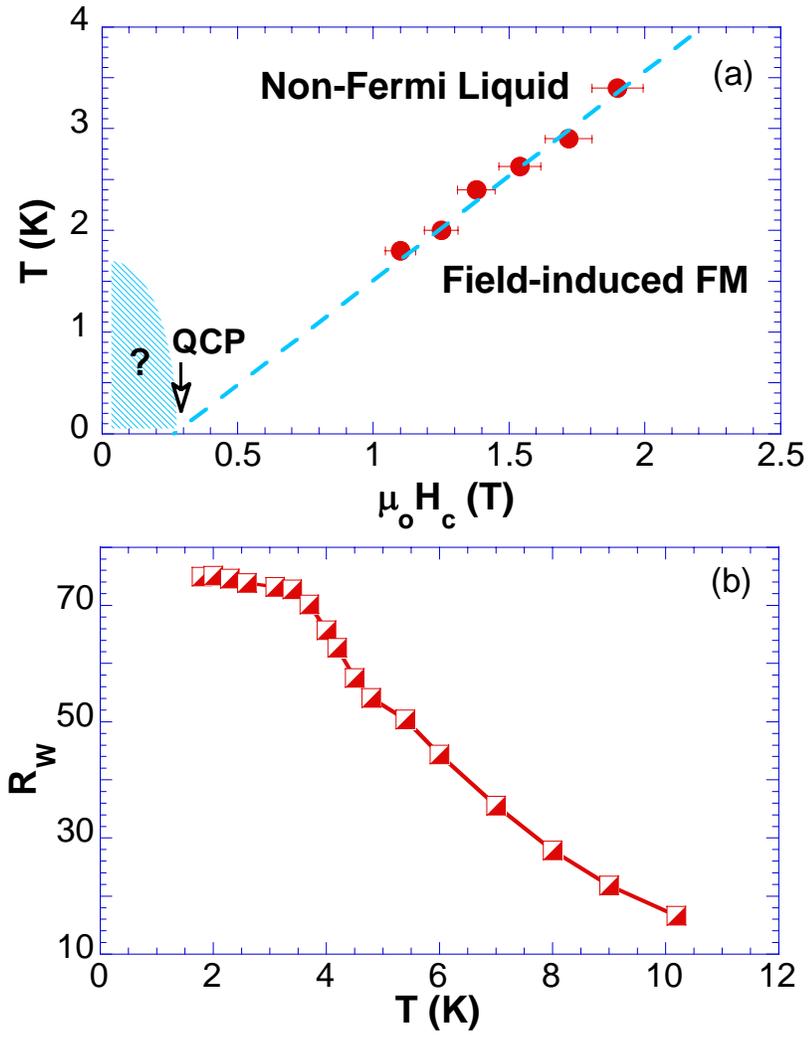

Fig. 3

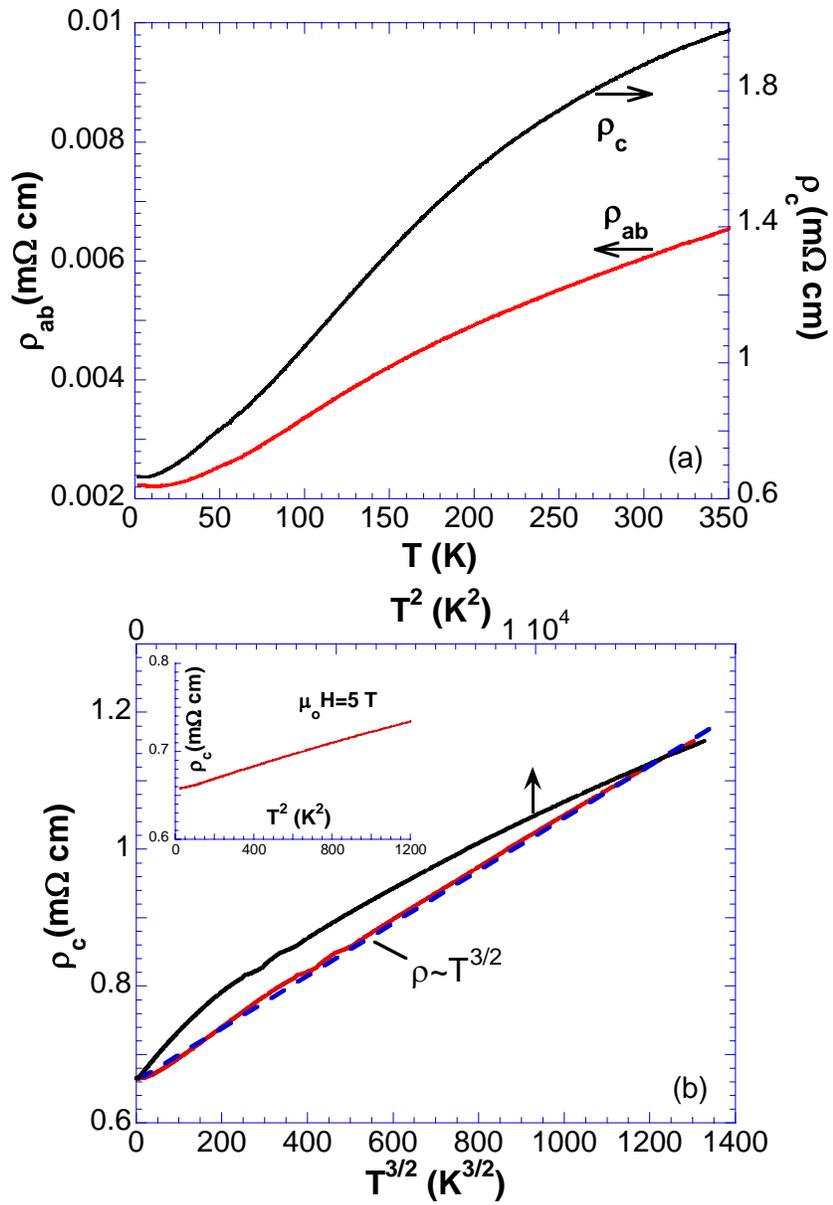

Fig. 4